\documentclass[journal,twoside,web]{ieeecolor2}
\usepackage{generic}
\usepackage{cite}
\usepackage{amsmath,amssymb,amsfonts}
\usepackage{algorithmic}
\usepackage{graphicx}
\usepackage{textcomp}
\usepackage{booktabs, makecell, url}

\def\BibTeX{{\rm B\kern-.05em{\sc i\kern-.025em b}\kern-.08em
    T\kern-.1667em\lower.7ex\hbox{E}\kern-.125emX}}
\begin{document}
\bstctlcite{IEEEtran:BSTcontrol}
\title{Real-Time Gradient Waveform Design for Arbitrary $k$-Space Trajectories}
\author{
Rui Luo, Hongzhang Huang, Qinfang Miao, Jian Xu, Peng Hu, Haikun Qi
\thanks{Manuscript submitted July 10, 2025. This work was supported in part by the High Technology Research and Development Center of the Ministry of Science and Technology of China under Grant SQ2022YFC2400133, in part by the Explorer Program of Shanghai Municipality under Grant 23TS1400300. (Corresponding author: Haikun Qi.)}
\thanks{Rui Luo is with the School of Biomedical Engineering \& State Key Laboratory of Advanced Medical Materials and Devices, ShanghaiTech University, Shanghai 201210, China (e-mail: \text{luorui2023@shanghaitech.edu.cn}).}
\thanks{Hongzhang Huang is with the School of Biomedical Engineering \& State Key Laboratory of Advanced Medical Materials and Devices, ShanghaiTech University, Shanghai 201210, China (e-mail: \text{huanghzh2023@shanghaitech.edu.cn}).}
\thanks{Qinfang Miao is with the School of Biomedical Engineering \& State Key Laboratory of Advanced Medical Materials and Devices, ShanghaiTech University, Shanghai 201210, China (e-mail: \text{miaoqf2022@shanghaitech.edu.cn}).}
\thanks{Jian Xu is with UIH America, Inc., Houston, TX 77054, USA (e-mail: \text{jian.xu01@united-imaging.com}).}
\thanks{Peng Hu is with the School of Biomedical Engineering \& State Key Laboratory of Advanced Medical Materials and Devices, ShanghaiTech University, Shanghai 201210, China (e-mail: \text{hupeng@shanghaitech.edu.cn}).}
\thanks{Haikun Qi is with the School of Biomedical Engineering \& State Key Laboratory of Advanced Medical Materials and Devices, ShanghaiTech University, Shanghai 201210, China (e-mail: \text{qihk@shanghaitech.edu.cn}).}
}

\maketitle

\begin{abstract}
\textbf{Objective: }To develop a real-time method for designing gradient waveforms for arbitrary $k$-space trajectories that are time-optimal and hardware-compliant.
\textbf{Methods: }The gradient waveform is solved recursively under both the slew-rate and the trajectory constraints. The gradient constraint is enforced by thresholding the $\ell_2$-norm of the next gradient vector. The constraints form a quadratic equation. To ensure the existence of the solution, a novel Discrete-Time Forward and Backward Sweep (DTFBS) strategy is proposed. To ensure the existence of the trajectory derivatives, the trajectory function is reparameterized as a piecewise cubic polynomial function with $C^2$ continuity. To ensure trajectory fidelity, the output gradient waveform is reparameterized by the finite difference of the trajectory samples. Simulation experiments across seven commonly adopted non-Cartesian trajectories were conducted to validate generality, time-optimality, real-time capability, slew-rate accuracy, and improvements over prior work. Imaging feasibility of the designed time-optimal gradient waveform was validated in phantom and in vivo experiments.
\textbf{Results: }The proposed method achieves a $>89\%$ reduction in computation time and simultaneously reduces slew-rate overshoot by $>98\%$ compared to the prior method across all involved trajectories. The computation time of the proposed method is shorter than the gradient duration for all tested cases, validating the real-time capability of the proposed method.
\textbf{Conclusions: }The proposed method enables real-time and hardware-compliant gradient waveform design, achieving significant reductions in computation time and slew-rate overshoot compared to the previous method.
\textbf{Significance: }This is the first method achieving real-time gradient waveform design for arbitrary $k$-space trajectories.
\end{abstract}


\section{Introduction}\label{sec:introduction}
\IEEEPARstart{O}{ne} of the significant advantages of non-Cartesian imaging \cite{ahnHighSpeedSpiralScanEcho1986} is its high sampling efficiency, which is crucial for Magnetic Resonance Imaging (MRI) due to the time-sensitive nature of the acquisition. More efficient trajectories \cite{gurneyDesignAnalysisPractical2006a,nollMultishotRosetteTrajectories1997,taoPartialFourierShells2019a,delattreSpiralDemystified2010,shuThreedimensionalMRIUndersampled2006,stobbeThreedimensionalYarnballKspace2021,speidelEfficient3DLowDiscrepancy2019} and faster reconstruction methods \cite{hamiltonLowrankDeepImage2023,fengGRASPProImProvingGRASP2020a,asslanderLowRankAlternating2018,zhaoLowRankMatrix2010} have been extensively investigated for non-Cartesian MRI. However, there are limited studies on accelerating gradient computation, which is a key step in non-Cartesian imaging. For 3D trajectories like Yarnball \cite{stobbeThreedimensionalYarnballKspace2021}, Shells \cite{taoPartialFourierShells2019a}, and Cones \cite{gurneyDesignAnalysisPractical2006a}, the current dominant method \cite{lustigFastMethodDesigning2008} takes up to several minutes for gradient computation before the scan, significantly slowing down the scan procedure. The slew-rate error also increases the risk of scan failure.

In earlier years, since only a few trajectories were available, non-Cartesian gradient waveform designs were tailored for specific trajectories \cite{salustriSimpleReliableSolutions1999,meyerRapidMethodOptimal2000,nollMultishotRosetteTrajectories1997}. All of these works rely on the mathematical form of the trajectory function and thus do not apply to arbitrary trajectories.

In 2008, Lustig \cite{lustigFastMethodDesigning2008} proposed to transplant optimal control theory \cite{kimComparisonPrincipleStateconstrained2005a} to non-Cartesian gradient waveform design because the $k$-space coordinate, gradient, and slew-rate have analogous relationships to kinematic concepts such as displacement, velocity, and acceleration. The optimal control theory solves the problem of minimizing the total traversal time through a given trajectory by planning its velocity with respect to arc-length. In practice, the input of the solver is an arc-length-uniformly sampled trajectory, and the output of the solver is the optimal velocity at the input arc-length samples. The arbitrarily parameterized trajectory function must be transformed to an arc-length-parameterized function by arc-length integration and numerical inversion, while the output velocity in the arc-length domain must be interpolated to the time domain before being passed to the gradient hardware. This method has been widely used in non-Cartesian imaging and is currently the dominant method \cite{speidelEfficient3DLowDiscrepancy2019,vannesjoImageReconstructionUsing2016,kasperMatchedfilterAcquisitionBOLD2014}.

The main drawback of Lustig's method comes from the interpolation in the last step: the density of the time-uniform sampling is much higher in the low-gradient regions, while the arc-length-uniform sampling is uniform throughout the whole trajectory. As a consequence, the slew-rate error is significantly higher in the low-gradient regions due to interpolation error. To compensate for this, the sampling density must be much higher than that of the gradient hardware.

Following Lustig's work, several derived methods were proposed \cite{chauffertProjectionAlgorithmGradient2016,davidsFastRobustDesign2015a}. However, these works involve iterative optimization and do not preserve the original trajectory shape. The computation time of previous methods ranges from several seconds to several minutes per trajectory, which is far behind the real-time requirement. There are also other works focused on gradient design \cite{loecherGradientOptimizationToolbox2020,pena-nogalesOptimizedDiffusionWeightingGradient2019}, which, however, are unrelated to non-Cartesian gradient waveforms. It is also notable that one recently published work \cite{pena-nogalesOptimizedDiffusionWeightingGradient2019} proposed a gradient optimization toolbox that targets the optimization of gradient waveforms for a variety of pulse sequences, but it is not targeted at readout gradients, while our work targets non-Cartesian readout gradients.

In this work, we propose an efficient, non-iterative method to design gradient waveforms for arbitrary $k$-space trajectories. The proposed method samples the gradient waveform with a constant time interval recursively, which aligns with the hardware-intrinsic sampling. The output gradient waveform is naturally time-parameterized and free of interpolation. We further propose several strategies to ensure the existence of solutions, the existence of derivatives, and trajectory fidelity, ensuring the proposed method is applicable to any trajectory under strict hardware constraints and preserves the original trajectory shape.

\section{Methods}\label{sec:Methods}
\subsection{From Trajectory to Slew-Rate}
In MRI, the $k$-space sampling position $\mathbf{k}(t)$ is given by

\begin{align}
    \mathbf{k}(t) = \gamma \int_{0}^{t} \mathbf{g}(\tau) d\tau,
\end{align}

where $\mathbf{g}(t)$ is the gradient vector and $\gamma$ is the gyromagnetic ratio, which is a constant for a given imaging nucleus. For notational simplicity, we absorb this constant into the gradient, yielding:

\begin{align}
    \mathbf{k}(t) = \int_{0}^{t} \mathbf{g}(\tau) d\tau,
    \label{eq:Grad}
\end{align}

The slew-rate $\mathbf{s}(t)$ is defined as

\begin{align}
    \mathbf{s}(t) = \dfrac{d\mathbf{g}(t)}{dt}.
    \label{eq:Slew}
\end{align}

For a given gradient coil, rotation-invariant hardware constraints apply:

\begin{align}
    \forall t:\quad \|\mathbf{g}(t)\| \le G_\mathrm{lim},\quad \|\mathbf{s}(t)\| \le S_\mathrm{lim},
    \label{eq:HardConstraint}
\end{align}

where $G_\mathrm{lim}$ and $S_\mathrm{lim}$ are the maximum allowed gradient and slew-rate magnitudes constrained by the hardware, respectively. From Eq.~\ref{eq:Grad} and Eq.~\ref{eq:Slew}, it can be seen that the gradient and slew-rate are equivalent to the velocity and acceleration of the $k$-space trajectory.

\subsection{Trajectory Definition}
Any $k$-space trajectory in MRI can be parameterized as

\begin{align}
    \mathbf{k}(p) = (k_x(p),\;k_y(p),\;k_z(p)),\quad p\in[P_0,P_1],
    \label{eq:GenTrajFunc}
\end{align}

where $p\in[P_0,P_1]$ is the trajectory parameter, and $k_x$, $k_y$, and $k_z$ are arbitrary scalar functions. For example, for a 2D Spiral trajectory, $p$ denotes the azimuthal angle $\theta$, $k_x(p)$ and $k_y(p)$ denote the coordinates of the Spiral trajectory, $k_z(p)$ is always zero, yielding

\begin{align}
    \mathbf{k}(p) = (A p \cos{p},\; A p \sin{p},\; 0),\quad p\in[0,\frac{1}{2A}].
    \label{eq:SpiralTrajFunc}
\end{align}

where $A$ is the shape parameter of the Spiral trajectory. Our goal is to find a gradient waveform $\mathbf{g}(t)$ such that its integral traces the arbitrary curve defined by Eq.~\ref{eq:GenTrajFunc}, while satisfying the hardware constraints in Eq.~\ref{eq:HardConstraint}.

\subsection{Constraints}\label{sec:constraint}
Given the gradient vector $\mathbf{g}(t_\mathrm{this})$ at a certain time point $t_\mathrm{this}$, the gradient vector $\mathbf{g}(t_\mathrm{next})$ at the next time point $t_\mathrm{next}$, and time step $\Delta t=t_\mathrm{next}-t_\mathrm{this}$, assuming the gradient field varies linearly between two temporal samples, the slew-rate magnitude within $(t_\mathrm{this},\,t_\mathrm{next}]$ is defined by $\|\mathbf{g}(t_\mathrm{next})-\mathbf{g}(t_\mathrm{this})\|/\Delta t$. The discrete slew-rate magnitude and the gradient magnitude at $t_\mathrm{next}$ should satisfy the hardware constraints in Eq.~\ref{eq:HardConstraint}:

\begin{align}
    \forall t_\mathrm{next}:&\quad \|\mathbf{g}(t_\mathrm{next})\| \le G_\mathrm{lim},\notag\\
    &\quad \|\mathbf{g}(t_\mathrm{next}) - \mathbf{g}(t_\mathrm{this})\| \le S_\mathrm{lim} \Delta t.
    \label{eq:HardConstraint2}
\end{align}

In addition to the hardware constraints, another constraint is that the next gradient $\mathbf{g}(t_\mathrm{next})$ should be parallel to the tangent at the next trajectory sample (hereafter referred to as the trajectory constraint):

\begin{align}
    \mathbf{g}(t_\mathrm{next}) \parallel \left. \dfrac{d\mathbf{k}}{dp} \right|_{t_\mathrm{next}}.
\end{align}

Since $\Delta t$ is sufficiently small, the tangent at the current trajectory sample is close to that at the next sample:

\begin{align}
    \dfrac{d\mathbf{k}(p_\mathrm{next})}{dp} \approx \dfrac{d\mathbf{k}(p_\mathrm{this})}{dp},
\end{align}

where $p_\mathrm{this}=p(t_\mathrm{this})$ and $p_\mathrm{next}=p(t_\mathrm{next})$, which yields:

\begin{align}
    \mathbf{g}(t_\mathrm{next}) \parallel \dfrac{d\mathbf{k}(p_\mathrm{this})}{dp}.
    \label{eq:DkConstraint}
\end{align}

\begin{figure}[t!]
    \centering
    \includegraphics{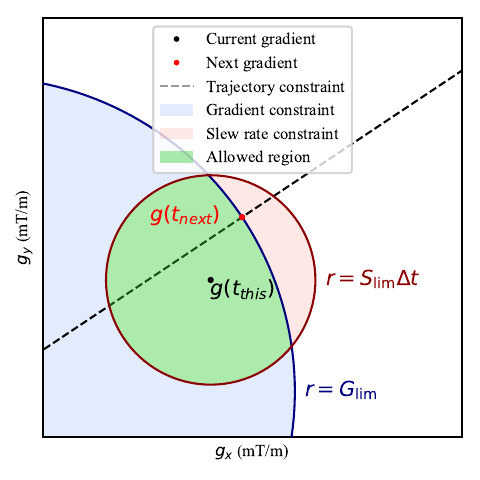}
    \caption{The permissible region (in green) for the next gradient vector is the intersection of the gradient constraint region $\|\mathbf{g}\| \le G_\mathrm{lim}$ (in blue) and the slew-rate constraint region $\|\mathbf{g} - \mathbf{g}(t_\mathrm{this})\| \le S_\mathrm{lim} \Delta t$ (in red). The dashed line denotes all possible gradient vectors parallel to $d\mathbf{k}(t_\mathrm{this})$.}
    \label{fig:GradSol}
\end{figure}

The hardware constraints defined by Eq.~\ref{eq:HardConstraint2} and the trajectory constraint defined by Eq.~\ref{eq:DkConstraint} are depicted as two Euclidean balls and a straight line in the gradient vector space as shown in Fig.~\ref{fig:GradSol}.

The given initial values $G_0$ and $G_1$ of gradient magnitude are also constraints, termed ``boundary constraints''. Additionally, the escape velocity also applies to the kinematic behavior of the $k$-space sample, which is referred to as the ``curvature constraint''. The curvature constraint will be explained in Section~\ref{sec:existsol}.

\subsection{Recursive Solution for Arbitrary Trajectories}\label{sec:recsol}
In this section, we present a recursive solution for arbitrary $k$-space trajectories to construct the gradient waveform under the hardware constraints. We solve the next gradient vector by solving its direction and magnitude separately. The direction of a gradient vector is denoted by the corresponding unit vector. Additionally, we use second-order Runge--Kutta (RK2) integration to approximate the next trajectory parameter.

As introduced in the previous subsection, the direction of the next gradient vector is parallel to $d\mathbf{k}(p_\mathrm{this})$, thus,

\begin{align}
    \hat{\mathbf{g}}(t_\mathrm{next}) = \dfrac{d\mathbf{k}(p_\mathrm{this})/dp}{\|d\mathbf{k}(p_\mathrm{this})/dp\|} \operatorname{sgn}\left(\dfrac{dp(t_\mathrm{this})}{dt}\right),
    \label{eq:GUnitSol}
\end{align}

where $\hat{\mathbf{g}}(t_\mathrm{next})$ denotes the unit vector of $\mathbf{g}(t_\mathrm{next})$.

Since a time-optimal gradient waveform must maximize either the gradient magnitude or the slew-rate magnitude \cite{lustigFastMethodDesigning2008}, the norm of the next gradient vector can be determined by solving for the farthest intersection between the slew-rate constraint $\|\mathbf{g} - \mathbf{g}(t_\mathrm{this})\| \le S_\mathrm{lim} \Delta t$ and the trajectory constraint $\mathbf{g}(t_\mathrm{next}) \parallel d\mathbf{k}(t_\mathrm{this})$, subject to the maximum gradient magnitude $G_\mathrm{lim}$, which is a simple geometrical problem: finding the intersection of the line defined by $\mathbf{g}(t_\mathrm{next}) \parallel d\mathbf{k}(t_\mathrm{this})$ and the Euclidean ball defined by $\|\mathbf{g} - \mathbf{g}(t_\mathrm{this})\| \le S_\mathrm{lim} \Delta t$, which forms a simple quadratic equation. The solution is given by:

\begin{align}
    \|\mathbf{g}(t_\mathrm{next})\| = \min\left\{Y,\, G_\mathrm{lim} \right\},
    \label{eq:GNormSol}
\end{align}

where

\begin{align}
    Y &= \dfrac{-b + \sqrt{b^2 - 4ac}}{2a}, \notag \\
    a &= 1, \notag \\
    b &= -2\langle \mathbf{g}(t_\mathrm{this}), \hat{\mathbf{g}}(t_\mathrm{next}) \rangle, \notag \\
    c &= \|\mathbf{g}(t_\mathrm{this})\|^2 - S_\mathrm{lim}^2 \Delta t^2. \notag
\end{align}

Note that $\langle \cdot, \cdot \rangle$ denotes the inner product of two vectors.

Combining the unit vector in Eq.~\ref{eq:GUnitSol} and magnitude in Eq.~\ref{eq:GNormSol}, the final solution for $\mathbf{g}(t_\mathrm{next})$ is:

\begin{align}
    \mathbf{g}(t_\mathrm{next}) = \hat{\mathbf{g}}(t_\mathrm{next})\, \|\mathbf{g}(t_\mathrm{next})\|.
\end{align}

The RK2 approximation for $p_\mathrm{next}$ is:

\begin{align}
    p_\mathrm{next} = p_\mathrm{this} + \dfrac{(k_1 + k_2) dl}{2},
\end{align}

where

\begin{align}
    dl &= \|\mathbf{g}(t_\mathrm{next})\| \Delta t, \notag \\
    k_1 &= \dfrac{1}{\|d\mathbf{k}(p_\mathrm{this})/dp\|}\operatorname{sgn}\left(\dfrac{dp}{dt}\right), \notag \\
    k_2 &= \dfrac{1}{\|d\mathbf{k}(p_\mathrm{this} + k_1 dl)/dp\|}\operatorname{sgn}\left(\dfrac{dp}{dt}\right). \notag
\end{align}

By running the above steps recursively, a gradient sequence $\{\mathbf{g}(t_i)\;|\;i=0,1,\ldots,N\}$ and a parameter sequence $\{p(t_i)\;|\;i=0,1,\ldots,N\}$ can be obtained simultaneously.

\subsection{Existence of Solution}\label{sec:existsol}
It is noted that the solution to Eq.~\ref{eq:GNormSol} does not always exist. In this section, we provide a method to ensure the existence of the solution.

Intuitively, it is noted that the gradient magnitude, i.e., the velocity of the $k$-space sample, will monotonically increase until $G_\mathrm{lim}$ while being recursively updated according to Eq.~\ref{eq:GNormSol}, as shown in Fig.~\ref{fig:GradSol}. However, once the velocity exceeds the escape velocity of the current trajectory curvature, it will be impossible to trace the given trajectory with the available acceleration, i.e., slew-rate.

To address this problem, we propose the Discrete-Time Forward and Backward Sweep strategy (DTFBS). Note that unlike the traditional Forward and Backward Sweep strategy \cite{kimComparisonPrincipleStateconstrained2005a}, DTFBS has a constant sampling interval in the time domain rather than in the arc-length domain or curve parameter domain. This property is crucial for highly efficient sampling as shown in Section~\ref{sec:Res_SampDenAna}.

DTFBS takes the minimum of the forward solution and the backward solution acquired in Section~\ref{sec:recsol}, denoted as:

\begin{align}
    \forall p:\quad \|\mathbf g^F_B(p)\|=\min\{\|\mathbf{g}_\mathrm{forw}(p)\|,\,\|\mathbf{g}_\mathrm{back}(p)\|\},
    \label{eq:DSweep}
\end{align}

where the forward solution $\mathbf{g}_\mathrm{forw}(p)$ is the gradient vector sequence solved by the recursive method in Section~\ref{sec:recsol}, but additionally throttled by the escape velocity $G_\mathrm{esc}$, as shown below:

\begin{align}
    \|\mathbf{g}_\mathrm{forw}(p)\| &= \min\left\{Y,\, G_\mathrm{lim},\, G_\mathrm{esc} \right\},\notag \\
    \textit{where}\quad&G_\mathrm{esc} = \sqrt{S_\mathrm{lim}/\kappa},\notag \\
    &\kappa = \|\dfrac{d\mathbf{k}}{dp} \times \dfrac{d^2\mathbf{k}}{dp^2}\|/\|\dfrac{d\mathbf{k}}{dp}\|^3.
\end{align}

The backward solution $\mathbf{g}_\mathrm{back}(p)$ is the gradient vector sequence solved by the recursive method in Section~\ref{sec:recsol}, but with reversed sampling direction along the trajectory and also throttled by $G_\mathrm{esc}$, as shown below:

\begin{align}
    \|\mathbf{g}_\mathrm{back}(p)\| = \min\left\{Y,\, G_\mathrm{lim},\, G_\mathrm{esc} \right\}.
\end{align}

The final solution is given by Eq.~\ref{eq:DSweep}. In practice, Eq.~\ref{eq:DSweep} can be embedded within the forward solution, while the backward solution is obtained in advance, given by:

\begin{align}
    \|\mathbf g^F_B(p)\|=\min\{Y,\|\mathbf{g}_\mathrm{back}(p)\|\}.
\end{align}

Since the sampling of $\mathbf{g}_\mathrm{back}(p)$ and $\mathbf{g}_\mathrm{forw}(p)$ is not aligned, an interpolation is needed when evaluating $\mathbf{g}_\mathrm{back}(p)$ in the forward sweep. In practice, a simple linear interpolation is enough without introducing significant error.

Intuitively, DTFBS ensures that, near regions of high curvature, the gradient magnitude in the forward sweep will decrease at the same slew-rate as in the backward sweep, i.e., at the maximum available slew-rate. Therefore, the DTFBS strategy guarantees that the gradient waveform not only operates at the maximum slew-rate for time-optimality, but also remains below the escape velocity--even at curvature maxima--ensuring that Eq.~\ref{eq:GNormSol} always has a solution. The time-optimality will be validated by comparing the gradient duration with that of the prior work in Section~\ref{sec:Res_PerfComp}.

Boundary constraints can be satisfied by setting the initial condition in the forward sweep to $\|\mathbf g(t_\mathrm{this})\|=G_0$ and the initial condition in the backward sweep to $\|\mathbf g(t_\mathrm{this})\|=G_1$.

\subsection{Existence of Derivatives}
As described in the previous section, the first- and second-order derivatives of the trajectory function must exist for calculating the gradient direction and the escape velocity. However, closed-form derivatives are not always available, while using finite differences is neither precise nor efficient. We therefore introduce a preprocessing step before the recursive process: reparameterizing the trajectory function into a piecewise cubic polynomial by cubic spline interpolation, which is referred to as trajectory reparameterization.

Cubic spline interpolation is performed to find a piecewise function $\tilde{f}(x)$ defined by a set of cubic polynomial functions $\{f_i(x)=a_ix^3+b_ix^2+c_ix+d_i,\,x\in[x_i,x_{i+1}]\;|\;i\in\{0,...,N-1\}\}$ that satisfies

\begin{align}
    \forall\quad i\in&\{0,...,N-2\},\;o\in\{1,2\}: \notag\\
    &f_i^{(o)}(x_{i+1})=f_{i+1}^{(o)}(x_{i+1}),
\end{align}

and
    
\begin{align}
    \forall\quad i\in&\{0,...,N-1\}: \notag\\
    &f_i(x_i)=y_i,\quad f_i(x_{i+1})=y_{i+1},
\end{align}

where $o$ denotes the order of derivatives, and $\{(x_i,y_i)\;|\;i\in\{0,...,N\}\}$ denotes the input point set.

According to the above constraints, the number of equations is $(4N-2)$ and the number of unknowns is $4N$. We can solve all the coefficients by assuming $f_0''(x_0)=0$ and $f_{N-1}''(x_N)=0$.

Thus, the found piecewise function $\tilde{f}(x)$ has the following properties:
\begin{enumerate}
    \item Passes exactly through the point set $\{(x_i,y_i)\;|\;i\in\{0,...,N\}\}$.
    \item $\tilde{f}$, $\tilde{f}'$, and $\tilde{f}''$ exist and are continuous on $[x_0, x_N]$ ($C^2$ continuity).
\end{enumerate}

Thus, by first sampling the original trajectory function $\mathbf k(p)$ to get $\{\mathbf k_i\;|\;i\in\{0,...,N\}\}$ and then fitting each axis by a piecewise cubic polynomial to get $\tilde{\mathbf k}(p)$, we can guarantee the existence of the first- and second-order derivatives (property 2) while preserving the shape of the trajectory (property 1) by replacing $\mathbf k(p)$ with $\tilde{\mathbf k}(p)$.

\subsection{Trajectory Fidelity}
Based on the recursive algorithm introduced in Section~\ref{sec:recsol}, we obtain two sequences after the solving process: $\mathbf{g}(t_i)$ and $p(t_i)$. Therefore, there are two approaches to derive the final gradient waveform:
\begin{enumerate}
    \item Use $\mathbf{g}(t_i)$ directly.
    \item Evaluate the trajectory function at $p(t_i)$ to obtain $\mathbf{k}(t_i)$, and then reparameterize $\mathbf{g}(t_i)$ by taking the finite difference of $\mathbf{k}(t_i)$ as shown below:
    \begin{align}
        \mathbf{g}(t_i) = (\mathbf{k}(t_{i+1})-\mathbf{k}(t_i))/\Delta t,
        \label{eq:GradRep}
    \end{align}
    which is referred to as gradient reparameterization.
\end{enumerate}
The first approach ensures the exact hardware constraints by construction, while the second approach ensures exact sampling accuracy, which is preferred over the first approach for trajectory fidelity. The preferred approach may lead to a trade-off in slew-rate accuracy, which can be compensated for by employing temporal oversampling by using a smaller time step than the hardware temporal resolution during the recursive solving process. The parameter sequence $p(t_i)$ is then downsampled before calculating the gradient waveform using Eq.~\ref{eq:GradRep}.

\subsection{Code Availability}
The implementation of the proposed method, as well as the scripts for reproducing the main results and the trajectory library for the imaging experiment, will be publicly available at \url{https://github.com/RyanShanghaitech/MrAutoGrad} upon publication of this study.

\section{Experiments}\label{sec:Experiment}
In this section, we describe the conducted experiments. We first compared the arc-length-uniform sampling used by the optimal control method \cite{lustigFastMethodDesigning2008} with the hardware-intrinsic time-uniform sampling to estimate the distribution of the interpolation error in the optimal control method. Afterwards, we validated the time complexity of our method and evaluated the influence of the temporal oversampling factor $R_\mathrm{samp}$ (defined by Eq.~\ref{eq:Rsamp}). Finally, we compared the performance metrics of the proposed method and the previous optimal control method, including the number of recursive steps $N_\mathrm{step}$, corresponding computation time $T_\mathrm{com}$, and slew-rate overshoot $E_\mathrm{slew}$ (defined by Eq.~\ref{eq:Eslew}). Real-time capability was determined by whether the computation time was shorter than the output gradient duration. We also illustrated the duration of the gradient waveform generated by the optimal control method and the proposed method to validate the time-optimality of the proposed method. To evaluate the imaging feasibility of the time-optimal gradient waveforms, we conducted both phantom and in vivo imaging experiments using the gradient waveforms generated by our method. Seven commonly adopted 2D and 3D non-Cartesian trajectories \cite{ahnHighSpeedSpiralScanEcho1986,delattreSpiralDemystified2010,nollMultishotRosetteTrajectories1997,stobbeThreedimensionalYarnballKspace2021,gurneyDesignAnalysisPractical2006a,speidelEfficient3DLowDiscrepancy2019} were included in the simulation and imaging experiments.

\subsection{Sampling Density Analysis}
In this experiment, we quantified the density of the arc-length-uniform sampling used in the optimal control method \cite{lustigFastMethodDesigning2008}, and the density of the time-uniform sampling used in our hardware-intrinsic method. By comparing the sampling density, the distribution of interpolation error in the previous method can be roughly estimated.

To quantify the sampling density, we used the reciprocal of the Euclidean distance between two adjacent samples in the $k$-space trajectory to define the sampling density, as shown below:

\begin{align}
    \rho_i=\frac{1}{\|\Delta \mathbf{k}_i\|}=\frac{1}{\|\mathbf{k}_{i+1}-\mathbf{k}_{i}\|}
    \label{eq:SampDen}
\end{align}

The sampling density maps of the arc-length-uniform sampling $\rho_\mathrm{arc}$ and time-uniform sampling $\rho_\mathrm{time}$ were quantified exemplarily on a Spiral trajectory. We sampled $N=473$ points with a constant arc-length interval and a constant time interval and calculated the sampling density using Eq.~\ref{eq:SampDen}.

\subsection{Time Complexity Analysis}
According to Section~\ref{sec:Methods}, the computation time of our method is proportional to the number of recursive steps, while the recursive steps are positively correlated with the duration of the gradient waveform. Thus, we hypothesize the proposed algorithm has $O(N)$ time complexity.

To validate this hypothesis, the proposed method was performed for 64 Spiral trajectories with different shapes. The duration of the gradient waveform $T_\mathrm{grad}$ and the corresponding computation time $T_\mathrm{com}$ were recorded. Then we fitted the data points with a linear function and calculated the coefficient of determination $R^2$, which is defined by:

\begin{align}
    R^2 = 1 - \frac{\sum_{i} (y_i - \hat{y}_i)^2}{\sum_{i} (y_i - \bar{y})^2},
\end{align}

where $y_i$ denotes the measured $T_\mathrm{com}$ of the $i$-th experiment, $\hat{y}_i$ denotes the predicted value from the linear fit of the $i$-th $T_\mathrm{grad}$, and $\bar{y}$ is the mean of the $T_\mathrm{com}$ of all the experiments. $R^2$ measures the proportion of the variance in the observed $T_\mathrm{com}$ explained by the linear fit, with $R^2=1$ indicating a perfect fit.

\subsection{Temporal Oversampling Analysis}
The gradient reparameterization strategy adopted to ensure trajectory accuracy introduces a tunable parameter, the temporal oversampling factor $R_\mathrm{samp}$, which controls the trade-off between slew-rate control accuracy and computational speed. $R_\mathrm{samp}$ is defined by:

\begin{align}
    R_\mathrm{samp}=\frac{\Delta t_\mathrm{hw}}{\Delta t},
    \label{eq:Rsamp}
\end{align}

where $\Delta t_\mathrm{hw}$ is the temporal resolution of the gradient hardware, and $\Delta t$ is the time step used in the recursive solver in Section~\ref{sec:recsol}.

In this experiment, a simulation was performed to evaluate the effect of this parameter for a representative Spiral trajectory.

\begin{table*}[t]
    \centering
    \caption{Summary of the imaging parameters for phantom and in vivo experiments.}
    \label{tab:ScanPara}
    \begin{tabular}{lccccccccc}
        \toprule
        Trajectory & \makecell{Res\\(mm)} & \makecell{FOV\\(mm)} & \makecell{FA\\(deg)} & \makecell{BW\\(Hz)} & \makecell{Grad\\(mT/m)} & \makecell{Slew\\(T/m/s)} & \makecell{TE\\(ms)} & \makecell{TR\\(ms)} & \makecell{Scan Time\\(s)} \\
        \midrule
        Spiral & 0.5 & 256 & 10 & -- & 50 & 50 & 1.25 & 14.01 & 2.86 \\
        VD. Spiral & 0.5 & 256 & 10 & -- & 50 & 50 & 1.25 & 12.73 & 5.19 \\
        Rosette & 0.5 & 256 & 10 & -- & 50 & 50 & (Mixed) & 20.35 & 6.53 \\
        Cartesian (2D) & 0.5 & 256 & 10 & 500 & -- & -- & 4.21 & 9.00 & 4.61 \\
        Yarnball & 1.0 & 256 & 10 & -- & 50 & 50 & 1.10 & 16.19 & 265.26 \\
        Cones & 1.0 & 256 & 10 & -- & 50 & 50 & 1.10 & 12.10 & 297.44 \\
        Seiffert Spiral & 1.0 & 256 & 10 & -- & 50 & 50 & 1.10 & 14.65 & 317.64 \\
        Stack-of-Spiral & 1.0 & 256 & 10 & -- & 50 & 50 & 1.10 & 12.10 & 198.25 \\
        Cartesian (3D) & 1.0 & 256 & 10 & 500 & -- & -- & 2.88 & 6.90 & 452.20 \\
        \bottomrule
        \multicolumn{10}{p{.7\textwidth}}{\footnotesize
        Res: resolution; FOV: field of view; FA: flip angle; BW: bandwidth; Grad: gradient magnitude limit; Slew: slew-rate magnitude limit; TE: echo time; TR: repetition time. ``--'' indicates not applicable.}
    \end{tabular}
\end{table*}

\begin{figure*}[t!]
    \centering
    \includegraphics{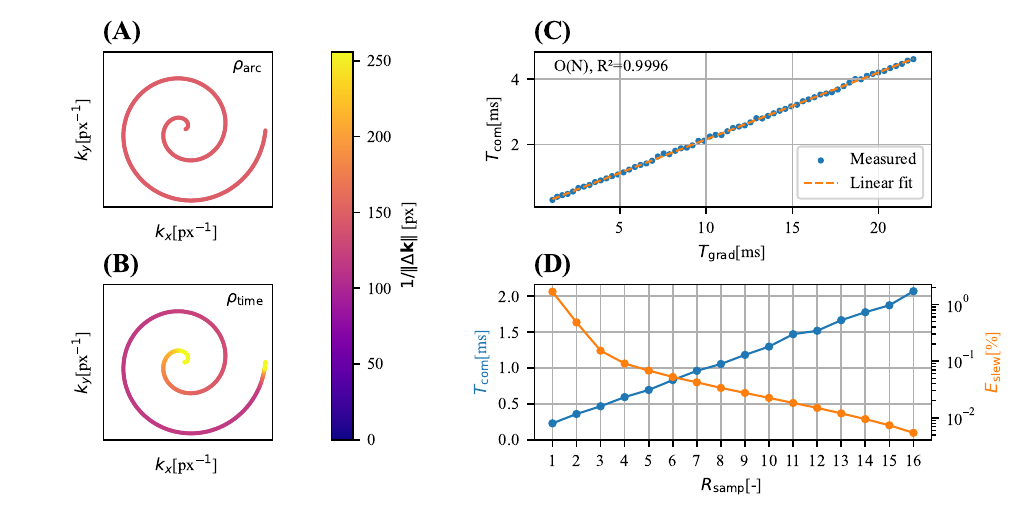}
    \caption{(A) Sampling density of arc-length-uniform sampling used in the optimal control method \cite{lustigFastMethodDesigning2008}. (B) Sampling density of time-uniform sampling, i.e., intrinsic sampling density of gradient hardware as was adopted in the proposed method. (C) Time complexity analysis showing the linear relationship between $T_\mathrm{com}$ and $T_\mathrm{grad}$ of our method. Each blue dot represents an independent experiment. The orange dashed line is fitted with the least-squares method. (D) The blue line represents the relationship between $R_\mathrm{samp}$ and $T_\mathrm{com}$. The orange line shows how $R_\mathrm{samp}$ influences $E_\mathrm{slew}$.}
    \label{fig:MultiFig}
\end{figure*}

\begin{figure*}[t!]
    \centering
    \includegraphics{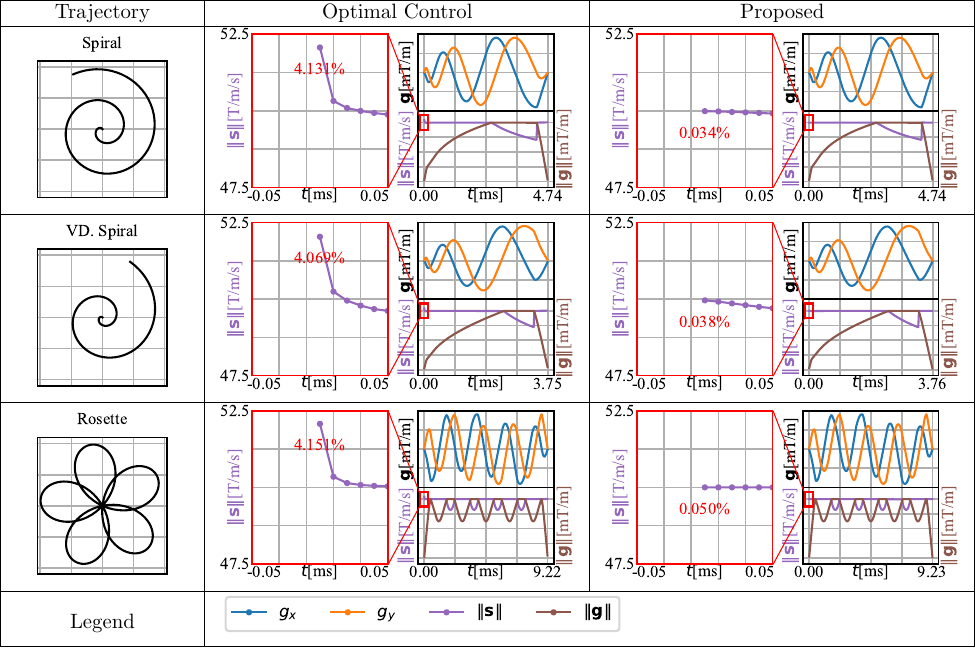}
    \caption{Visualization of the gradient waveforms and the corresponding gradient magnitude and slew-rate magnitude generated by the optimal control method and the proposed method for various 2D non-Cartesian trajectories. The initial part of the slew-rate curve is zoomed in to compare the slew-rate overshoot percentages between the optimal control method and the proposed method.}
    \label{fig:wavetab2D}
\end{figure*}

\subsection{Performance Comparison}
The proposed method was compared with the widely adopted optimal control method \cite{lustigFastMethodDesigning2008} for designing time-optimal gradient waveforms for a range of non-Cartesian trajectories including Spiral \cite{ahnHighSpeedSpiralScanEcho1986}, Variable Density Spiral \cite{delattreSpiralDemystified2010} (VD. Spiral), Rosette \cite{nollMultishotRosetteTrajectories1997}, Yarnball \cite{stobbeThreedimensionalYarnballKspace2021}, Cones \cite{gurneyDesignAnalysisPractical2006a}, Seiffert Spiral \cite{speidelEfficient3DLowDiscrepancy2019}, and Stack-of-Spiral. Both the proposed method and the previous optimal control method were implemented in C++. The simulation was conducted on a 4.90 GHz CPU (Intel® Core™ i7-12700). The code of the optimal control method was adapted from the official implementation\footnote{Refer to the ``Time Optimal Gradient Design'' section at \url{https://people.eecs.berkeley.edu/~mlustig/Software.html}} with minimal modifications for compatibility.

During the simulation, the slew-rate magnitude was limited to 50~$\mathrm{T/m/s}$ and the gradient magnitude was limited to 20~$\mathrm{mT/m}$, which can be adjusted according to specific hardware constraints. Each trajectory was simulated 100 times to calculate the mean and standard deviation (std.) of the computation time ($T_\mathrm{com}$). For evaluating the slew-rate control accuracy, the slew-rate overshoot ($E_\mathrm{slew}$) was calculated as follows:

\begin{align}
    E_\mathrm{slew}=\frac{\max_i{\{s[i]\}}-S_\mathrm{lim}}{S_\mathrm{lim}}\times100\%
    \label{eq:Eslew}
\end{align}

\subsection{Imaging Experiments}
To validate the feasibility of the time-optimal gradient waveforms generated by the proposed method for actual MRI, phantom and in vivo imaging experiments were conducted by imaging the NIST/ISMRM phantom (System Phantom, Model 130 from CaliberMRI\textregistered) and the knee of a healthy subject after IRB approval and informed consent. A non-Cartesian FLASH imaging protocol using the proposed method was implemented in a 3.0 T United Imaging MR scanner. The slew-rate magnitude limit was consistent with that used in the simulation, while the gradient magnitude limit was relaxed to 50~$\mathrm{mT/m}$ to increase the imaging speed. The non-Cartesian MRI reconstruction was performed using non-uniform fast Fourier transform (NUFFT) \cite{barnettAliasingErrorExp$vsqrt1z^2$2020,barnettParallelNonuniformFast2019}, where the sampling density compensation function was calculated using a numerical iterative method \cite{pipeSamplingDensityCompensation1999a}. Imaging parameters are summarized in Table~\ref{tab:ScanPara}. It is noted that a 1-3-3-1 water-excitation pulse was adopted to suppress the fat signal for all the non-Cartesian imaging sequences to reduce the off-resonance artifacts. However, in Cartesian imaging the number of excitations can be significantly larger than in non-Cartesian imaging. Considering that the water-excitation pulse significantly extends the repetition time, it was not adopted for Cartesian imaging to avoid prohibitively long scan duration.

\section{Results}\label{sec:Results}
\subsection{Sampling Density Analysis}\label{sec:Res_SampDenAna}
The sampling densities of arc-length-uniform sampling and time-uniform sampling are compared in Fig.~\ref{fig:MultiFig}A, B, respectively. The density of arc-length-uniform sampling approaches a constant along the Spiral trajectory (from $147.1$px to $147.8$px) because sampling is uniform in arc-length by construction. However, the density of the proposed time-uniform sampling ranges from $117.440$px to $3455.055$px, with high sampling density in the beginning and the end of the trajectory. Thus, when performing interpolation from the arc-length domain to the time domain in the previous optimal control method, the interpolation error can be expected at the beginning and the end of the gradient waveforms.

\begin{table*}[t!]
    \centering
    \caption{Benchmark between the optimal control method and the proposed method}
    \label{tab:Benchmark}
    \begin{tabular}{l cccc cccc ccc}
        \toprule
        & \multicolumn{4}{c}{Optimal Control} & \multicolumn{4}{c}{Proposed} & \multicolumn{3}{c}{Improvement} \\
        \cmidrule(lr){2-5} \cmidrule(lr){6-9} \cmidrule(lr){10-12}
        \makecell{Trajectory} &
        \makecell{$N_\mathrm{step}$} &
        \makecell{$T_\mathrm{com}$\\{[ms]}} &
        \makecell{$T_\mathrm{grad}$\\{[ms]}} &
        \makecell{$E_\mathrm{slew}$\\{[\%]}} &
        \makecell{$N_\mathrm{step}$} &
        \makecell{$T_\mathrm{com}$\\{[ms]}} &
        \makecell{$T_\mathrm{grad}$\\{[ms]}} &
        \makecell{$E_\mathrm{slew}$\\{[\%]}} &
        \makecell{$N_\mathrm{step}$\\Reduction} &
        \makecell{$T_\mathrm{com}$\\Reduction} &
        \makecell{$E_\mathrm{slew}$\\Reduction} \\
        \midrule
        Spiral           & 90632  & 11.00$\pm$0.39 & 4.73 & 4.131 &  7555 & 1.05$\pm$0.05 & 4.73 & 0.034 & 91.66\% & 90.42\% & 99.18\% \\
        VD. Spiral       & 65528  & 8.21$\pm$0.21  & 3.74 & 4.069 &  5990 & 0.85$\pm$0.02 & 3.75 & 0.038 & 90.86\% & 89.67\% & 99.08\% \\
        Rosette          & 179868 & 22.44$\pm$0.83 & 9.21 & 4.151 & 14572 & 1.95$\pm$0.06 & 9.22 & 0.050 & 91.90\% & 91.33\% & 98.80\% \\
        Yarnball         & 191202 & 22.32$\pm$0.98 & 9.11 & 4.148 & 14544 & 1.97$\pm$0.17 & 9.11 & 0.032 & 92.39\% & 91.15\% & 99.24\% \\
        Cones            & 79000  & 9.94$\pm$0.34  & 4.32 & 4.136 &  6891 & 1.05$\pm$0.15 & 4.32 & 0.036 & 91.28\% & 89.48\% & 99.13\% \\
        Seiffert Spiral  & 142398 & 18.59$\pm$0.47 & 7.32 & 4.099 & 11706 & 1.62$\pm$0.16 & 7.32 & 0.038 & 91.78\% & 91.27\% & 99.08\% \\
        Stack-of-Spiral  & 90632  & 11.07$\pm$0.59 & 4.73 & 4.131 &  7555 & 1.09$\pm$0.09 & 4.73 & 0.034 & 91.66\% & 90.14\% & 99.18\% \\
        \bottomrule
        \multicolumn{12}{p{0.90\textwidth}}{\footnotesize
        $N_\mathrm{step}$: number of recursive steps; $T_\mathrm{com}$: computation time (mean$\pm$std); $T_\mathrm{grad}$: duration of readout gradient; $E_\mathrm{slew}$: slew-rate overshoot. Real-time capability is guaranteed when $T_\mathrm{com}<T_\mathrm{grad}$.}
    \end{tabular}
\end{table*}

\begin{figure*}[t!]
    \centering
    \includegraphics{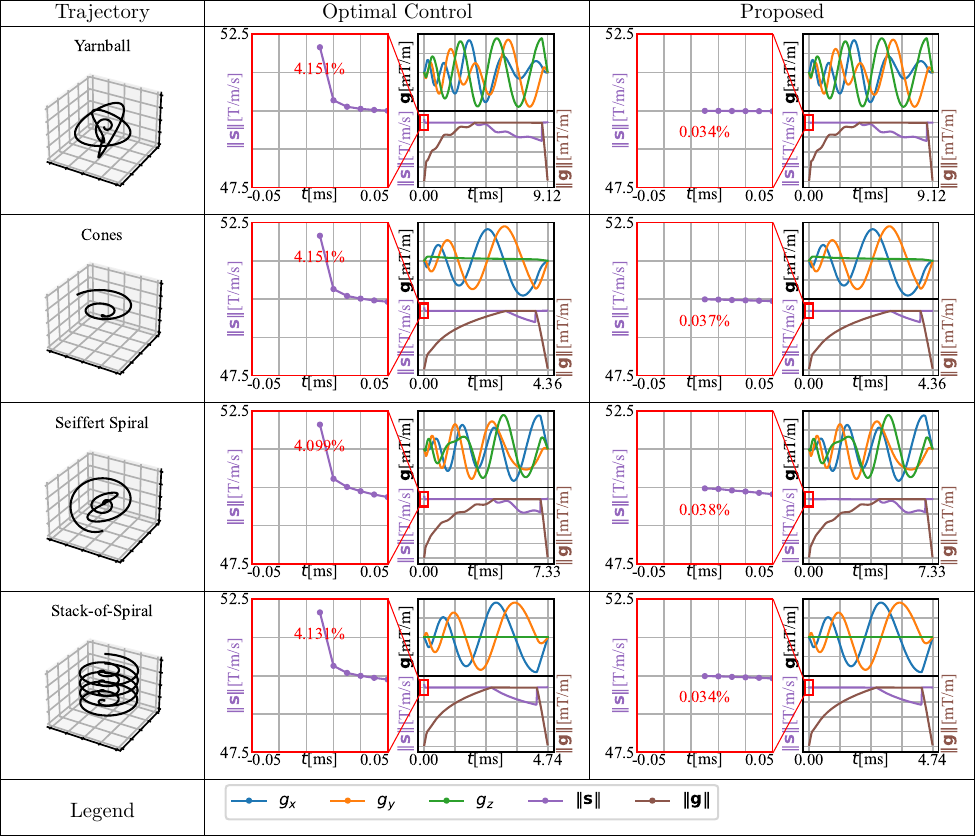}
    \caption{Visualization of the gradient waveforms and the corresponding gradient magnitude and slew-rate magnitude generated by the optimal control method and the proposed method for various 3D non-Cartesian trajectories. The initial part of the slew-rate curve is zoomed in to compare the slew-rate overshoot percentages between the optimal control method and the proposed method.}
    \label{fig:wavetab3D}
\end{figure*}

\subsection{Time Complexity Analysis}
The relationship between $T_\mathrm{com}$ and $T_\mathrm{grad}$ is illustrated in Fig.~\ref{fig:MultiFig}C. It can be seen that the computation time is linearly proportional to the duration of the output gradient. The coefficient of determination $R^2$ for the linear fitting is $0.9996$, strongly validating the time complexity of $O(N)$.

\subsection{Temporal Oversampling Analysis}\label{sec:Res_TemOvAna}
The influence of the temporal oversampling factor $R_\mathrm{samp}$ on the computation speed and slew-rate control accuracy for the proposed method is demonstrated in Fig.~\ref{fig:MultiFig}D. It can be seen that without temporal oversampling, slew-rate overshoot is near $1\%$. Otherwise, $E_\mathrm{slew}$ decreases exponentially with increasing $R_\mathrm{samp}$ with $R_\mathrm{samp}>4$ achieving less than $0.1\%$ slew-rate overshoot, while $T_\mathrm{com}$ increases linearly with increasing $R_\mathrm{samp}$. To guarantee negligible slew-rate overshoot, a temporal oversampling factor of $8$ was adopted for the proposed method in the following experiments.

\subsection{Performance Comparison}\label{sec:Res_PerfComp}
Table~\ref{tab:Benchmark} summarizes the number of recursive steps $N_\mathrm{step}$, computation time $T_\mathrm{com}$, duration of the generated gradient waveform $T_\mathrm{grad}$, slew-rate overshoot $E_\mathrm{slew}$ of the optimal control method and the proposed method. The reduction percentages of computation time and slew-rate overshoot with the proposed method compared to the optimal control method are also included in Table~\ref{tab:Benchmark}. For all test trajectories, the optimal control method requires computation time longer than the duration of the gradient waveform, precluding its application for real-time gradient waveform design. In contrast, the proposed method uses nearly 90\% less time to generate the gradient waveforms, and its computation time is always shorter than the duration of the generated gradient waveform, facilitating real-time gradient computations. The sampling density maps in Fig.~\ref{fig:MultiFig} reveal the reason why our method is more efficient: to achieve acceptable accuracy, the arc-length-uniform sampling used by the optimal control method needs a relatively high sampling density across the whole $k$-space trajectory, leading to inefficient computation. Also, the inadequate sampling density at the beginning of the trajectory tends to yield slew-rate overshoot. The time-uniform sampling adopted in the proposed method enables trajectory-adaptive sampling density, thus, requiring much fewer recursive calculation steps and less computation time, and achieving better slew-rate control.

Fig.~\ref{fig:wavetab2D} and Fig.~\ref{fig:wavetab3D} visualize the designed gradient waveform of the optimal control method and the proposed method, encompassing different 2D and 3D non-Cartesian trajectories. It can be seen that the optimal control method tends to have slew-rate overshoot at the beginning of the trajectory, which coincides with the simulation results in Section~\ref{sec:Res_SampDenAna}, while the proposed method achieves close-to-zero slew-rate overshoot for all simulated trajectories. The durations of the generated gradient waveforms are similar between the optimal control method and the proposed method, validating the time-optimality of the proposed method. 

\subsection{Imaging Experiments}
\begin{figure}[t!]
    \centering
    \includegraphics{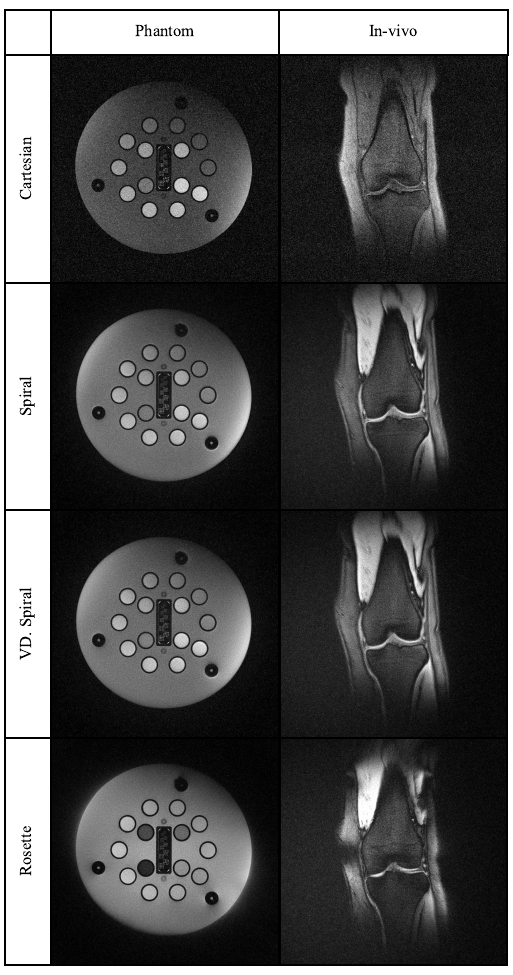}
    \caption{The phantom and in vivo images acquired with three 2D non-Cartesian trajectories implemented using the proposed method in comparison with the images acquired with Cartesian sampling. The data acquired with the Rosette trajectory have multiple echoes, and averaging the multi-echo data may cause signal cancellation for certain tissues, leading to dark regions in the Rosette image. }
    \label{fig:Result2D}
\end{figure}

\begin{figure*}[t!]
    \centering
    \includegraphics{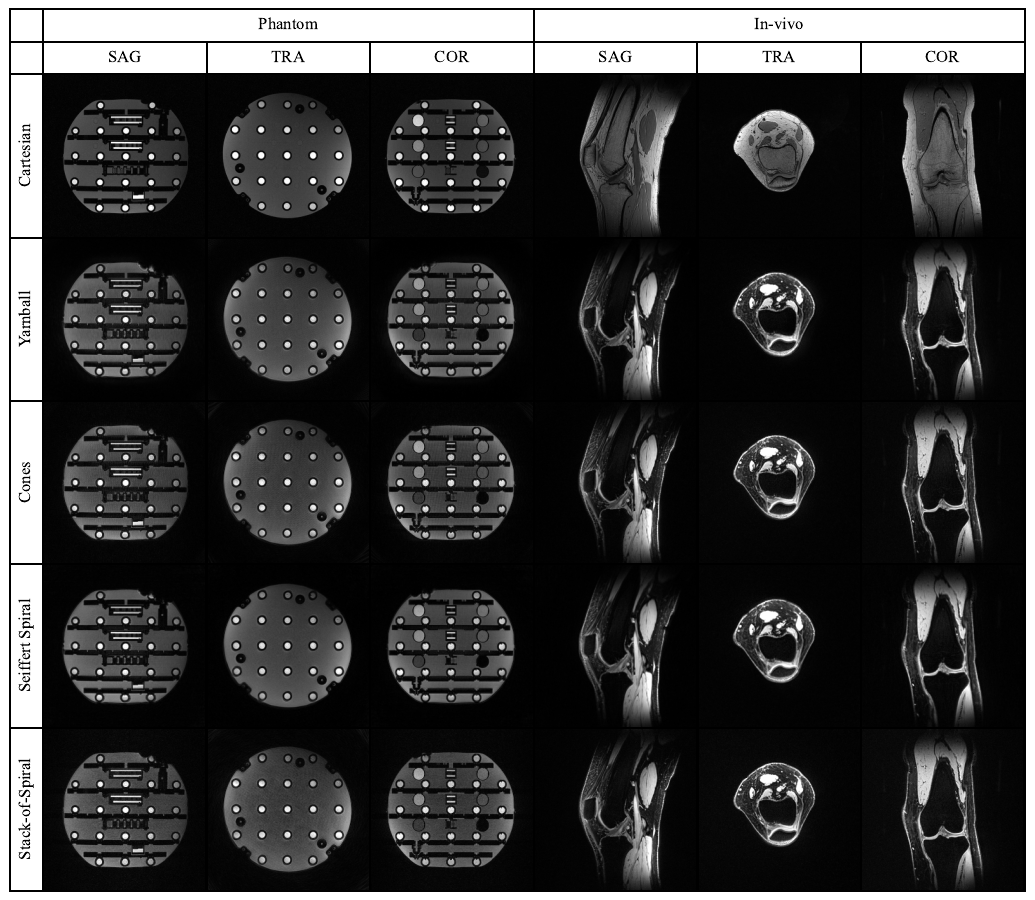}
    \caption{The phantom and in vivo images acquired with four 3D non-Cartesian trajectories implemented using the proposed method in comparison with the images acquired with Cartesian sampling. SAG: sagittal; TRA: transverse; COR: coronal.}
    \label{fig:Result3D}
\end{figure*}

The 2D and 3D imaging results of the phantom and human knee are respectively presented in Fig.~\ref{fig:Result2D} and Fig.~\ref{fig:Result3D}. All the non-Cartesian acquisitions with the readout gradients designed by the proposed method were successfully performed, generating images depicting clear anatomical structures and free of distortions in comparison with the reference Cartesian acquisition.

\section{Discussion}\label{sec:Discussion}
In this study, an efficient gradient waveform design method was proposed. The proposed method contains a general-purpose recursive solver and several approaches to guarantee the existence of the solution, the existence of the derivatives, and trajectory fidelity. This method efficiently discretizes the gradient waveform in a hardware-intrinsic way and enables real-time gradient waveform design for arbitrary $k$-space trajectories.

A previous Spiral gradient design method proposed by Meyer et al. \cite{meyerRapidMethodOptimal2000} also benefits from visualizing the gradient vector in a Euclidean space. However, the basic assumption and the scope of our work are distinct from those of \cite{meyerRapidMethodOptimal2000}. In the previous work, since the trajectory is assumed to be Spiral, most of the derivation steps are fundamentally different, for example, when updating trajectory parameters, gradient vectors are integrated to obtain the $k$-space coordinate $\mathbf k$, from which the current radius is calculated by $\rho=\|\mathbf k\|$, and the azimuthal angle $\theta$ is then derived leveraging the proportionality of Spiral's analytical form $\rho=A\theta$. In our method, the trajectory parameter is updated by calculating its increment. The gradient magnitude $\|\mathbf g\|$ is first assumed to be constant, and then the increment in $k$-space $\|\Delta \mathbf k\|$ is calculated by multiplying the gradient magnitude $\|\mathbf g\|$ by the constant time step $\Delta t$. The parameter increment $\Delta p$ is then calculated by dividing the $k$-space distance $\|\Delta \mathbf k\|$ by the $\ell_2$-norm of the trajectory derivative $\|d\mathbf{k}/dp\|$, ensuring the generality for arbitrary trajectories. Besides, the efforts to ensure the existence of the solution, the existence of the derivatives, and the trajectory fidelity are also new in our method and necessary for arbitrary trajectories.

In the context of automatic control, the traditional Forward and Backward Sweep (FBS) strategy is implemented by first fixing some control points along a curve, typically selected by constant arc-length interval or constant curve parameter interval, and then solving the state and co-state ordinary differential equations (ODEs) in the forward sweep and the backward sweep respectively \cite{kimComparisonPrincipleStateconstrained2005a}. In our proposed DTFBS, the control points are determined during runtime by using a constant time interval. A linear interpolation is needed to evaluate the backward solution in the forward sweep. This is crucial to guarantee highly efficient, hardware-intrinsic sampling for the gradient waveform design problem which saves nearly 90\% of recursive steps and computation time and reduces nearly 99\% slew-rate overshoot. Besides, forward and backward solutions are solved by quadratic equations rather than state and co-state ODEs. To the best of our knowledge, these characteristics have never appeared in prior work.

The proposed real-time method not only improves efficiency, but also may enable several new applications. Currently, all available 3D trajectories have a limited set of interleaf shapes, because when designing gradient waveforms with a non-real-time method such as the optimal control method, it would take huge amounts of time for designing a unique gradient for each interleaf. However, with the proposed real-time gradient waveform design, gradient computation is moved from the prescan stage to the scan stage, and no time is wasted on gradient precomputation. This allows highly efficient sampling trajectories which may have a unique shape for every interleaf, such as extending the Traveling Salesman Problem (TSP) trajectory \cite{sharmaKSpaceTrajectoryDesign2020} to 3D. Another application is the adaptive non-Cartesian imaging, in which real-time gradient design allows the trajectories to be adjusted on the fly. For example, in cardiac MRI, the $k$-space data are collected only during the end-diastolic window to avoid cardiac motion \cite{rajiahCardiacMRIState2023}. When the heart rate rises, this acquisition window shortens and the pre-planned sampling pattern may no longer fit. By adjusting the trajectory or hardware constraints and recomputing the gradient waveforms in real-time, an acquisition with shortened window may be enabled to be compatible with increasing heart rates. When examining young children, silent imaging can be achieved by limiting the slew-rate \cite{zhouAcousticNoiseReduction2023}. Then, real-time gradient computation provides a way to tune the slew-rate during the exam according to the conditions of the scanned subject.

In general, real-time gradient waveform design provides exceptional flexibility for non-Cartesian MRI, allowing for dynamically adjusting the sampling pattern and hardware constraints to handle unexpected events in clinical scans. It also provides new possibilities for trajectory design and optimization.

\section{Conclusion}\label{sec:Conclusion}
In this paper, we propose a non-Cartesian gradient waveform design method. This method is real-time capable, general-purpose, and hardware-compliant. Compared with the currently dominant optimal control method, the proposed method achieves a significant reduction in computation time and slew-rate overshoot. This is the first method capable of designing gradient waveforms for arbitrary $k$-space trajectories in real-time.

The proposed method is applicable to all non-Cartesian imaging trajectories. For general-purpose non-Cartesian imaging, this method improves scan efficiency by eliminating the non-Cartesian gradient precomputation. For adaptive non-Cartesian imaging, the method offers dynamic trajectories and adaptive hardware constraints to handle unexpected events, offering new possibilities for pulse sequence design. For trajectory design and optimization, this method enables the design of a unique, arbitrary shape for each interleaf, which substantially benefits the flexibility of trajectory design and optimization in non-Cartesian MRI.

\bibliographystyle{IEEEtran}
\bibliography{references}

\begin{thebibliography}{10}
\providecommand{\url}[1]{#1}
\csname url@samestyle\endcsname
\providecommand{\newblock}{\relax}
\providecommand{\bibinfo}[2]{#2}
\providecommand{\BIBentrySTDinterwordspacing}{\spaceskip=0pt\relax}
\providecommand{\BIBentryALTinterwordstretchfactor}{4}
\providecommand{\BIBentryALTinterwordspacing}{\spaceskip=\fontdimen2\font plus
\BIBentryALTinterwordstretchfactor\fontdimen3\font minus
  \fontdimen4\font\relax}
\providecommand{\BIBforeignlanguage}[2]{{%
\expandafter\ifx\csname l@#1\endcsname\relax
\typeout{** WARNING: IEEEtran.bst: No hyphenation pattern has been}%
\typeout{** loaded for the language `#1'. Using the pattern for}%
\typeout{** the default language instead.}%
\else
\language=\csname l@#1\endcsname
\fi
#2}}
\providecommand{\BIBdecl}{\relax}
\BIBdecl
\renewcommand{\BIBentryALTinterwordstretchfactor}{4}

\bibitem{ahnHighSpeedSpiralScanEcho1986}
C.~B. Ahn \emph{et~al.}, ``High-speed spiral-scan echo planar nmr imaging-i,''
  \emph{IEEE Trans. Med. Imag.}, vol.~5, no.~1, pp. 2--7, Mar. 1986.

\bibitem{gurneyDesignAnalysisPractical2006a}
P.~T. Gurney \emph{et~al.}, ``Design and analysis of a practical 3d cones
  trajectory,'' \emph{Magn Reson Med}, vol.~55, no.~3, pp. 575--582, Mar. 2006.

\bibitem{nollMultishotRosetteTrajectories1997}
D.~C. Noll, ``Multishot rosette trajectories for spectrally selective mr
  imaging,'' \emph{IEEE Trans. Med. Imag.}, vol.~16, no.~4, pp. 372--377, Aug.
  1997.

\bibitem{taoPartialFourierShells2019a}
S.~Tao \emph{et~al.}, ``Partial fourier shells trajectory for non-cartesian
  mri,'' \emph{Phys. Med. Biol.}, vol.~64, no.~4, Feb. 2019.

\bibitem{delattreSpiralDemystified2010}
B.~M.~A. Delattre \emph{et~al.}, ``Spiral demystified,'' \emph{Magn Reson
  Imaging}, vol.~28, no.~6, pp. 862--881, Jul. 2010.

\bibitem{shuThreedimensionalMRIUndersampled2006}
Y.~Shu \emph{et~al.}, ``Three-dimensional mri with an undersampled spherical
  shells trajectory,'' \emph{Magn Reson Med}, vol.~56, no.~3, pp. 553--562,
  Sep. 2006.

\bibitem{stobbeThreedimensionalYarnballKspace2021}
R.~W. Stobbe and C.~Beaulieu, ``Three-dimensional yarnball k-space acquisition
  for accelerated mri,'' \emph{Magn Reson Med}, vol.~85, no.~4, pp. 1840--1854,
  Apr. 2021.

\bibitem{speidelEfficient3DLowDiscrepancy2019}
T.~Speidel \emph{et~al.}, ``Efficient 3d low-discrepancy k-space sampling using
  highly adaptable seiffert spirals,'' \emph{IEEE Trans. Med. Imag.}, vol.~38,
  no.~8, pp. 1833--1840, Aug. 2019.

\bibitem{hamiltonLowrankDeepImage2023}
J.~I. Hamilton \emph{et~al.}, ``A low-rank deep image prior reconstruction for
  free-breathing ungated spiral functional cmr at 0.55 t and 1.5 t,''
  \emph{Magn Reson Mater Phys Biol Med}, vol.~36, no.~3, pp. 451--464, Apr.
  2023.

\bibitem{fengGRASPProImProvingGRASP2020a}
L.~Feng \emph{et~al.}, ``Grasp-pro: improving grasp dce-mri through
  self-calibrating subspace-modeling and contrast phase automation,''
  \emph{Magn Reson Med}, vol.~83, no.~1, pp. 94--108, Jan. 2020.

\bibitem{asslanderLowRankAlternating2018}
J.~Assländer \emph{et~al.}, ``Low rank alternating direction method of
  multipliers reconstruction for mr fingerprinting,'' \emph{Magn Reson Med},
  vol.~79, no.~1, pp. 83--96, Jan. 2018.

\bibitem{zhaoLowRankMatrix2010}
B.~Zhao \emph{et~al.}, ``Low rank matrix recovery for real-time cardiac mri,''
  in \emph{2010 IEEE International Symposium on Biomedical Imaging: From Nano
  to Macro}, Apr. 2010, pp. 996--999.

\bibitem{lustigFastMethodDesigning2008}
M.~Lustig \emph{et~al.}, ``A fast method for designing time-optimal gradient
  waveforms for arbitrary k-space trajectories,'' \emph{IEEE Trans. Med.
  Imag.}, vol.~27, no.~6, pp. 866--873, Jun. 2008.

\bibitem{salustriSimpleReliableSolutions1999}
C.~Salustri \emph{et~al.}, ``Simple but reliable solutions for spiral mri
  gradient design,'' \emph{J Magn Reson}, vol. 140, no.~2, pp. 347--350, Oct.
  1999.

\bibitem{meyerRapidMethodOptimal2000}
C.~H. Meyer and J.~M. Pauly, ``Rapid method of optimal gradient waveform design
  for mri,'' US Patent US6\,020\,739, Feb., 2000.

\bibitem{kimComparisonPrincipleStateconstrained2005a}
S.-J. Kim \emph{et~al.}, ``A comparison principle for state-constrained
  differential inequalities and its application to time-optimal control,''
  \emph{IEEE Trans. Autom. Control}, vol.~50, no.~7, pp. 967--983, Jul. 2005.

\bibitem{vannesjoImageReconstructionUsing2016}
S.~J. Vannesjo \emph{et~al.}, ``Image reconstruction using a gradient impulse
  response model for trajectory prediction,'' \emph{Magn Reson Med}, vol.~76,
  no.~1, pp. 45--58, Jul. 2016.

\bibitem{kasperMatchedfilterAcquisitionBOLD2014}
L.~Kasper \emph{et~al.}, ``Matched-filter acquisition for bold fmri,''
  \emph{NeuroImage}, vol. 100, pp. 145--160, Oct. 2014.

\bibitem{chauffertProjectionAlgorithmGradient2016}
N.~Chauffert \emph{et~al.}, ``A projection algorithm for gradient waveforms
  design in magnetic resonance imaging,'' \emph{IEEE Trans. Med. Imag.},
  vol.~35, no.~9, pp. 2026--2039, Sep. 2016.

\bibitem{davidsFastRobustDesign2015a}
M.~Davids \emph{et~al.}, ``Fast and robust design of time-optimal k-space
  trajectories in mri,'' \emph{IEEE Trans. Med. Imag.}, vol.~34, no.~2, pp.
  564--577, Feb. 2015.

\bibitem{loecherGradientOptimizationToolbox2020}
M.~Loecher \emph{et~al.}, ``A gradient optimization toolbox for general purpose
  time-optimal mri gradient waveform design,'' \emph{Magn Reson Med}, vol.~84,
  no.~6, pp. 3234--3245, Dec. 2020.

\bibitem{pena-nogalesOptimizedDiffusionWeightingGradient2019}
O.~Pe{~n}a-Nogales \emph{et~al.}, ``Optimized diffusion-weighting gradient
  waveform design (odgd) formulation for motion compensation and concomitant
  gradient nulling,'' \emph{Magn Reson Med}, vol.~81, no.~2, pp. 989--1003,
  Feb. 2019.

\bibitem{barnettAliasingErrorExp$vsqrt1z^2$2020}
A.~H. Barnett, ``Aliasing error of the exp($\beta\sqrt{1-z^2}$) kernel in the
  nonuniform fast fourier transform,'' Oct. 2020, arXiv:2001.09405.

\bibitem{barnettParallelNonuniformFast2019}
A.~H. Barnett \emph{et~al.}, ``A parallel non-uniform fast fourier transform
  library based on an "exponential of semicircle" kernel,'' Apr. 2019,
  arXiv:1808.06736.

\bibitem{pipeSamplingDensityCompensation1999a}
J.~G. Pipe and P.~Menon, ``Sampling density compensation in mri: Rationale and
  an iterative numerical solution,'' \emph{Magn Reson Med}, vol.~41, no.~1, pp.
  179--186, Jan. 1999.

\bibitem{sharmaKSpaceTrajectoryDesign2020}
S.~Sharma \emph{et~al.}, ``K-space trajectory design for reduced mri scan
  time,'' in \emph{ICASSP 2020 - 2020 IEEE International Conference on
  Acoustics, Speech and Signal Processing (ICASSP)}, May 2020, pp. 1120--1124.

\bibitem{rajiahCardiacMRIState2023}
P.~S. Rajiah \emph{et~al.}, ``Cardiac mri: State of the art,''
  \emph{Radiology}, vol. 307, no.~3, May 2023.

\bibitem{zhouAcousticNoiseReduction2023}
Z.~Zhou \emph{et~al.}, ``Acoustic noise reduction for spiral mri by gradient
  derating,'' \emph{Magn Reson Med}, vol.~90, no.~4, pp. 1547--1554, Oct. 2023.

\end{thebibliography}

\end{document}